\documentclass[12pt]{article}
\usepackage{epsfig}

\def\beq{\begin{equation}}
\def\eeq#1{\label{#1}\end{equation}}
\def\eeqn{\end{equation}}

\def\beqa{\begin{eqnarray}}
\def\eeqa#1{\label{#1}\end{eqnarray}}
\def\eeqan{\end{eqnarray}}

\let\bar=\overbar

\def\tr{{\mbox{\rm tr}}}

\def\Dslash{\not{\hbox{\kern-4pt $D$}}}
\def\dslash{\not{\hbox{\kern-2pt $\del$}}}

\def\msb{{\bar{\ssstyle M \kern -1pt S}}}

\usepackage{fancyhdr,graphicx}
\def\Title#1{\begin{center} {\Large {\bf #1} } \end{center}}

\begin{document}

\Title{Modeling  Hybrid Stars in Quark-Hadron Approaches}

\bigskip\bigskip


\begin{raggedright}

{\it S. Schramm$^1$, V. Dexheimer$^2$, R. Negreiros$^3$, T. Sch\"urhoff$^1$, J. Steinheimer$^4$ \\
\bigskip
$^1$FIAS, Ruth-Moufang-Str. 1, D-60438 Frankfurt am Main, Germany\\
$^2$Instituto de Fisica, Universidade Federal Fluminense, Nideroi, Brazil\\
$^3$Department of Physics, Kent State University, Kent OH 44242 USA\\
$^4$Lawrence Berkeley National Laboratory, Berkeley, CA 94720, USA}\\
\vspace{10pt}
{\tt Email: schramm@fias.uni-frankfurt.de}
\bigskip\bigskip
\end{raggedright}

\section{Introduction}

The study of neutron stars, or more general compact stars, is a topic of central interest in nuclear astrophysics.
Furthermore, neutron stars serve as the only physical systems whose properties can be used to infer information on cold and dense matter at several times nuclear saturation
density. Therefore, neutron star physics is ideally suited to complement the studies of ultra-relativistic heavy-ion collisions that sample
strongly interacting matter at high temperature and relatively small net baryon density. 

In general, in order to pin down or at least constrain the properties of dense matter, accurate measurements of neutron star properties  like masses, radii, rotational frequency, and cooling behavior are needed.
Here, in relatively recent times the reliable mass determination of the pulsar PSR J1614-2230 of $M = 1.97 \pm 0.04 M_\odot$ \cite{197} has introduced an important benchmark for modeling stars and strongly interacting matter. 
It puts constraints on the structure of compact stars and possible exotic phases in the core of the stars  as will be discussed in this article.
In order to investigate this point we will consider a model for star matter that includes hyperonic and quark degrees of freedom, and present results for compact star properties in the following.

\section{Hadronic Model Approach}

In our approach we combine hadronic and quark degrees of freedom in a single model, wherein all those degrees of freedom interact via mesonic fields.
This or some similar form of integrated description of hadrons and quarks is necessary for a realistic phase structure of strongly interacting matter.
In contrast to calculations that combine separate models for hadrons and quarks, which introduces a first-order transition between the two phases, here one can describe not only first-order phase transitions between hadrons and quarks but also second-order and cross-over transitions.
This is necessary for modeling
hadronic and quark regimes over the whole range of temperatures and densities, since we know from lattice QCD results that at low chemical potential the transition is a cross-over.

Our studies are based on a hadronic flavor-SU(3) model that includes the lowest SU(3) multiplets for baryons and mesons. A detailed discussion of this ansatz can
be found in \cite{Papazoglou:1997uw,Papazoglou:1998vr}.  
Assuming a linear coupling of mesonic fields and baryons, the various coupling constants can be determined by the three independent SU(3)-invariant terms with their corresponding
parameters $g_1, g_8,$ and $\alpha$:
\begin{eqnarray}
L_{BW} &=& - \sqrt{2} g_8 \left( \alpha \left[ BOBW\right]_F + (1 - \alpha) \left[ BOBW\right]_D \right)
\nonumber \\
& &- g_1 / \sqrt{3}\, \tr (BOB) \tr(W)
\end{eqnarray}
with the $F$ and $D$-type couplings 
\begin{eqnarray}
\left[ BOBW\right]_F ~&=&~ \tr (\overline{B}OBW - \overline{B}OWB)~~ \nonumber \\
\left[ BOBW\right]_D ~&=&~ \tr (\overline{B}OBW + \overline{B}OWB)~~.
\label{su3scheme}
\end{eqnarray}
Here, $B$ and $W$ are the SU(3) baryon and meson multiplets. The Dirac matrix $O$ depends on the quantum numbers of the specific mesonic multiplet,
i.e. in the case of a scalar coupling $O$ is the unit matrix and for vector mesons $O = \gamma^\mu$.

The baryon-vector meson interaction (0th component), which is the dominant interaction term at high densities, then reads
\begin{equation}
L_{Int}=-\sum_i \bar{\psi_i}[\gamma_0(g_{i\omega}\omega+g_{i\phi}\phi+g_{i\rho}\tau_3\rho)+m_i^*]\psi_i ,
\end{equation}
summing over the baryons $i$. The term includes the interaction with the non-strange and strange vector mesons $\omega, \rho$ and $\phi$.
The various couplings are related to $g_1, g_8$, and $\alpha$ via Eq. (\ref{su3scheme}). The effective baryon masses $m_i^*$  are defined by
the expression
\begin{equation}
m_i^* = g_{i\sigma} \sigma + g_{i\zeta} \zeta + g_{i\delta} \delta + \delta m_i ~~,
\label{efmas}
\end{equation}
with the scalar non-strange isoscalar field $\sigma$, isovector field $\delta$ as well as the scalar field with hidden strangeness
$\zeta$. There is also a small explicit mass term $\delta m_i$.
In addition, there are self-interaction terms for the mesons that lead to non-vanishing vacuum expectation values of 
the scalar fields. These in turn generate the masses of the baryons through Eq. (\ref{efmas}), and create an attractive potential in matter.
This hadronic model has been successfully applied in nuclear structure, heavy-ion, as well as nuclear astrophysics calculations \cite{Steinheimer:2007iy,Schramm:2002xi,Dexheimer:2008ax}.

\section{SU(6) and SU(3) symmetries}

Taking into account baryonic particles beyond nucleons in the description of the neutron star, especially hyperons, is a logical extension of a  standard SU(2) description of nuclear matter. A problem often encountered in calculations of hyper stars is that the additional degrees of freedom lead to a substantial softening of the equation of state and subsequently to small star masses well below the 2 solar mass value (see, e.g. \cite{arXiv:1006.5660,Schulze:2010zz,Agrawal:2012rx}). However, hyperons in neutron stars cannot just be disregarded, as from measurements of hyper nuclei it is known that at least the $\Lambda$ baryons are bound by about 30 MeV at saturation density, and therefore hyperons should be part of a comprehensive model.


As a commonly used guiding principle for choosing the coupling parameters of the various hyperons to the vector mesons a useful and natural assumption is the choice of SU(6) symmetry of flavor and spin following the ideas of universality and vector meson dominance.
In this limit there is only one invariant coupling of the vector meson fields to the baryonic $[56]$ multiplet (which also contains the spin 3/2 baryonic decuplet).
Therefore, the general SU(3) coupling scheme Eq.~(\ref{su3scheme}) with the three parameters $g_1, g_8, $ and $\alpha$ is simplified by the SU(6) relations between the parameters:
\begin{equation}
\alpha = 1 ~~~,~~~g_1/g_8 = \sqrt{6}~~~.
\label{su6}
\end{equation}
This leads to the following couplings of the baryons to the $\omega$ meson:
\begin{equation}
g_{N\omega} = -3 g_8~~,~~g_{\Lambda\omega} = g_{\Sigma\omega} = -2 g_8~~,~~g_{\Xi\omega} = - g_8~~
\label{gno}
\end{equation}
and to the $\phi$ meson
\begin{equation}
g_{N\phi} = 0~~,~~g_{\Lambda\phi} = g_{\Sigma\phi} = -\sqrt{2} g_8~~,~~g_{\Xi\phi} = - 2 \sqrt{2} g_8~~~.
\label{gnphi}
\end{equation}
Eqs.\,(\ref{gno}, \ref{gnphi}) correspond to simple quark counting rules for the coupling.
At high densities the repulsive interaction dominates over the scalar attraction. Increasing the singlet coupling strength $g_1$ (by keeping $g_{N\omega}$ constant) leads to increased repulsion in the hyperon sector, which in turn suppresses the hyperon fraction. This then has the effect of a relative stiffening of the equation of state and larger star masses as was discussed, e.g. in \cite{arXiv:1112.0234}. On the other hand, however, there has been an extensive experimental effort to determine the strange quark contribution to the vector form factor of the nucleon in parity-violating electron-nucleus scattering experiments (see. \cite {Armstrong:2012bi} for a review). The results clearly show that the strangeness contribution is very small (below 3 percent and consistent with 0). Theoretical modeling shows a natural direct link between the nucleonic vector strangeness content and the nucleon-$\phi$ coupling, which also turns out to be very small \cite{Schramm:1995qx}.  Therefore, there is strong evidence for a small coupling of the $\phi$ meson to the nucleon. However, following the SU(3) symmetric couplings in Eq. (\ref{su3scheme}) , the general expression for $g_{N\phi}$ is
\begin{equation}
g_{N\phi} = \frac{\sqrt{2}}{3} (4 \alpha - 1) g_8 - \frac{1}{\sqrt{3}} g_1~~~,
\end{equation}
which is zero for the choice Eq.\,(\ref{su6}), but becomes large for bigger deviations of $g_1$ from its SU(6) value. 
Therefore, breaking SU(6) but keeping SU(3) might generate larger star masses well beyond 2 solar masses, but it will lead to inconsistencies with the experimental findings.
A similar situation occurs with regard to the baryon 3/2 multiplet. In refs. \cite{Dexheimer:2008ax,Schurhoff:2010ph} we extended the model by including the baryonic spin 3/2 decuplet. Because of the mass of those particles  only the $\Delta$ baryons  can potentially occur in compact stars. A natural coupling choice of the $\Delta$ baryon to the omega field would be to a assume the same value as for the nucleons, in accordance with SU(6). We have studied the impact of changing the $\Delta$ coupling parameter to smaller values in ref. \cite{Schurhoff:2010ph}. The results show that a deviation of this value by more than 10 percent would reduce the maximum star mass below phenomenological values. This finding, of course, only holds in our specific chiral model. However, as the $\Delta$ baryons contain a large number of spin-isospin states (n=16), a significantly reduced vector repulsion for the Delta would most likely lead to a strong softening of the equation of state in any approach (similar results have been found in the analysis of quasi-elastic electron-nucleus scattering \cite{Wehrberger:1993zu}). Therefore, also in the case of the decuplet the choice of SU(6) seems reasonable. 

\section{Hadronic Stars}
\begin{figure}[th]
\centerline{\includegraphics[width=8cm]{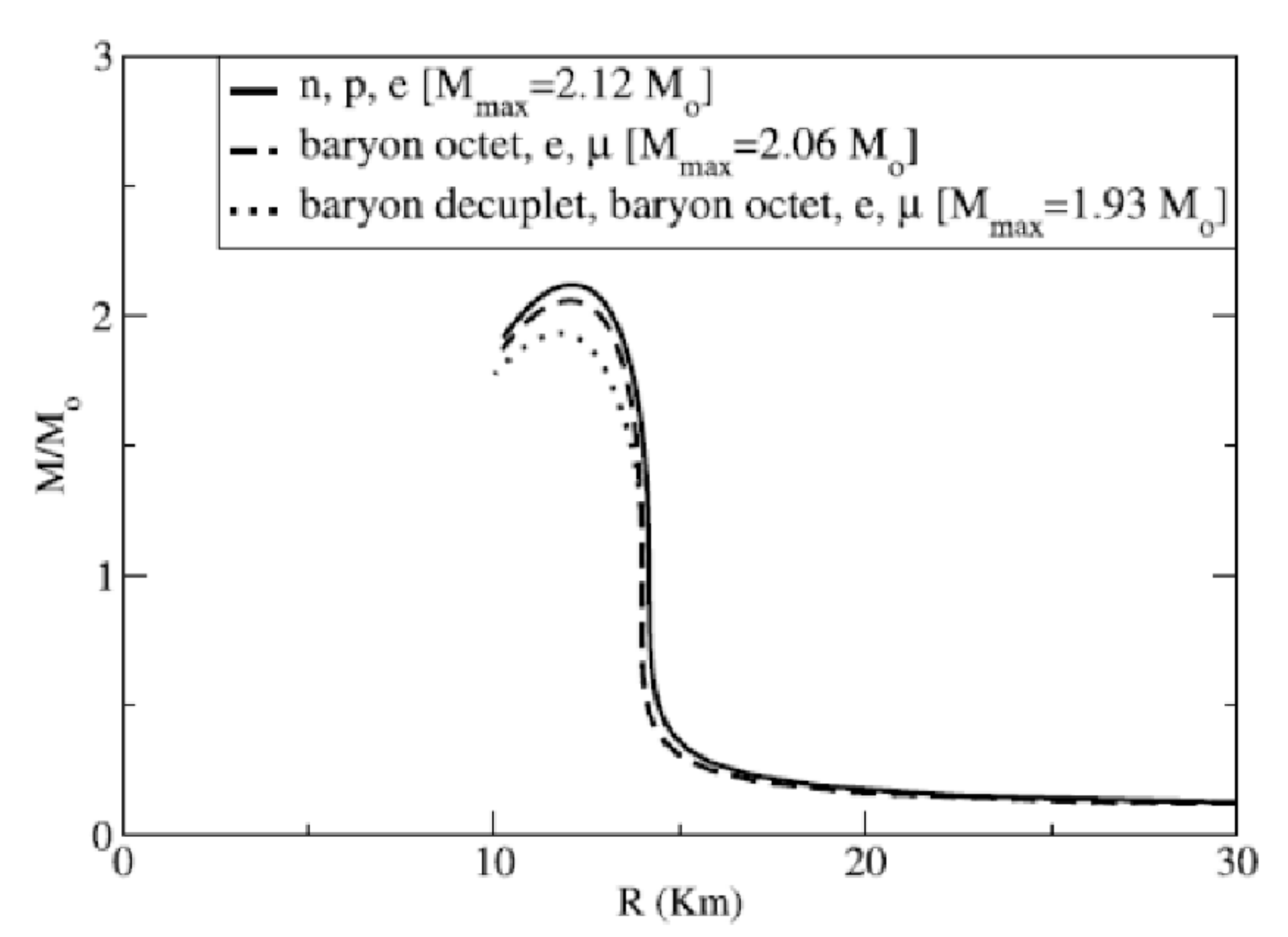}}
\vspace*{5pt}
\caption{Mass-radius diagram for compact stars with different baryonic degrees of freedom. Results for a star including nucleons, nucleons and hyperons, and in addition the
spin 3/2 decuplet are shown, respectively \protect\cite{Dexheimer:2008ax}.}
\label{mr}
\end{figure}

\begin{figure}[th]
\centerline{\includegraphics[width=8cm]{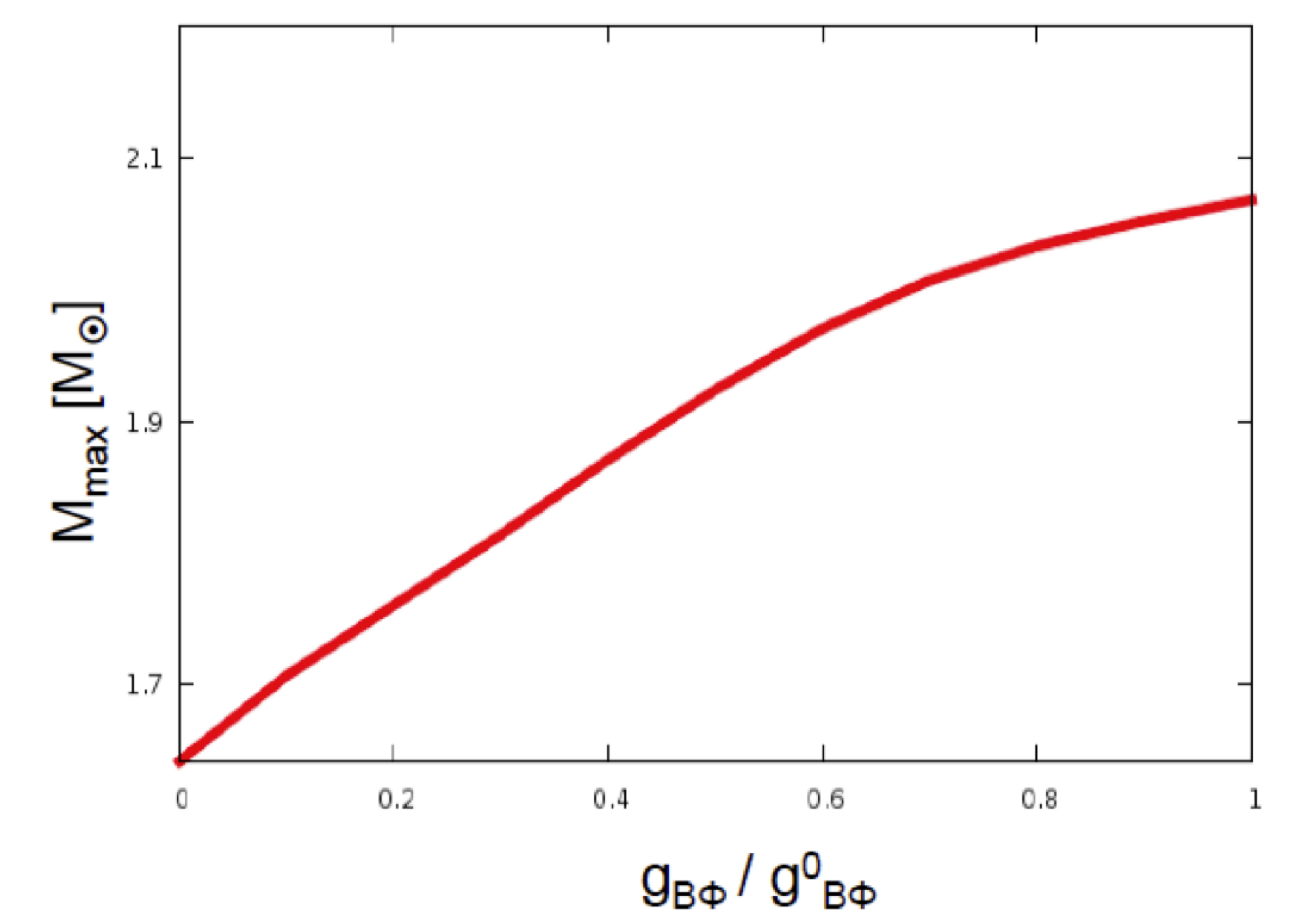}}
\vspace*{5pt}
\caption{Maximum mass of a hyper star as function of the coupling strength (normalized) of the hyperons to the $\phi$ vector field. Turning off the coupling leads to a drastic lowering of the mass.}
\label{phi}
\end{figure}

In our hadronic model, using  SU(6) couplings we determine masses and radii of compact stars by solving the Tolman-Oppenheimer-Volkov
equations for a static and spherically symmetric star in the usual manner \cite{tov1,tov2}. The results (Fig. \ref{mr}) show maximum star masses up to 2.12 solar masses for a purely nucleonic star, and a maximum mass of 2.06 M$_\odot$ including hyperons \cite{Dexheimer:2008ax}. 
Thus, the addition of hyperons does not result in a significant lowering of the star masses. A major reason for this behavior is illustrated in Fig. \ref{phi}. The figure shows the maximum star mass as function of the coupling strength of the hyperons to the $\phi$ meson relative to the value used for Fig. \ref{mr}. Reducing the coupling, one can observe a significant drop of the mass to a value of about 1.65 M$\odot$ in the case of vanishing coupling. This clearly illustrates that model descriptions, which include hyperons but do not take into account the corresponding repulsive effects due to strange meson exchange, will have problems with a strong softening of their equation of state of star matter, and in consequence obtain significantly reduced star masses. However, the occurrence of the repulsive $\phi$ meson field is a natural consequence of any flavor SU(3) model.

A different way of implementing chiral symmetry in an effective hadronic Lagrangian is based on the so-called parity doublet model. Within this approach the baryonic degrees of freedom are
extended to doublets ($B^+,B^-$) with relative opposite parity. 
Within such an approach, chiral symmetry restoration results in the onset of degeneracy of both parity partners. 
The details of the approach are outlined in \cite{Steinheimer:2011ea,Dexheimer:2012eu}.
\begin{figure}[]
\centerline{\includegraphics[width=8.4cm]{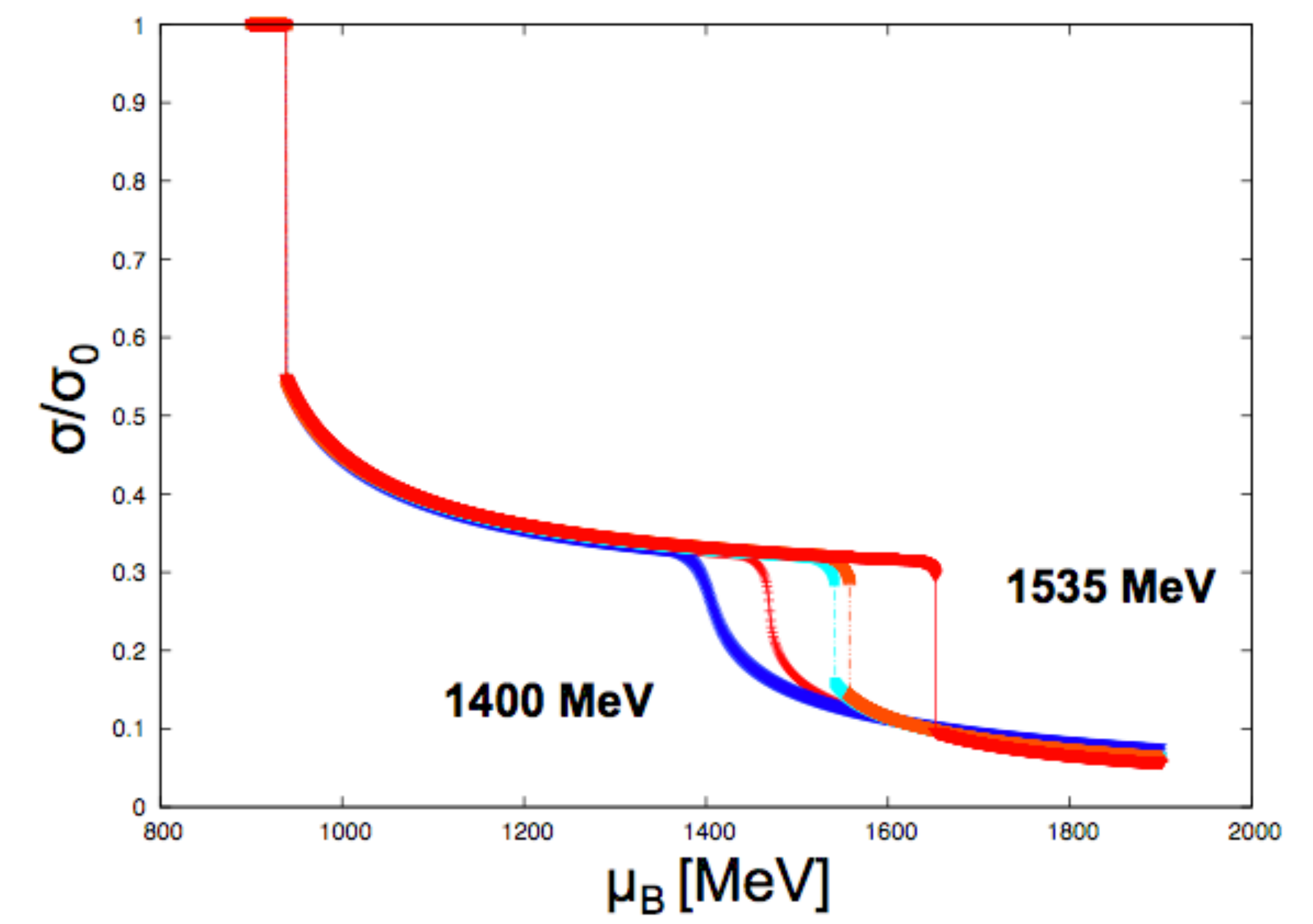}}
\vspace*{5pt}
\caption{Behavior of the scalar field $\sigma$ normalized to its vacuum value. 
Results for different parameter sets corresponding to different masses of the parity partner of the nucleon are shown.
A switch from a cross-over phase transition to a first-order transition can be seen for masses larger than about 1470 MeV.}
\label{sigmamnpar}
\end{figure}
Effectively, the basic difference of this approach is a modified expression for the effective baryonic mass, which for isospin symmetric matter reads
\begin{equation}
m^*_i(\pm) = \sqrt{ \left[ (g_{\sigma i} \sigma + g_{\zeta i}  \zeta)^2 + (m_0+n_s m_s)^2 \right]}
\mp g'_{\sigma i} \sigma \mp g'_{\zeta i} \zeta~.
 \label{mef}
\end{equation}
with different masses for the two parity states $\pm$ . The term $n_s m_s$ represents the explicit mass due to the strange quark content of the respective baryonic state. The parity-doublet models allows for a chirally-invariant mass term $m_0$
as the parity violation of the mass term from one component of the doublet is compensated by the other one, in the same way as it is the case for the chirally invariant combination $\sigma^2 + \pi^2$ of the scalar and pseudo scalar fields in linear $\sigma$ models. The doublet formulation also introduces a second set of coupling constants $g'$ that generates the mass splitting.
From expression Eq. (\ref{mef}) one can see that for vanishing scalar fields the states are degenerate.
In the case of the nucleon a likely candidate for such an opposite-parity state is the N(1535) resonance. There are some states which might be the corresponding counterparts in the hyperon sector (\cite{Steinheimer:2011ea}), but this question has not been settled. For simplicity we assume the same mass gap for all baryons and their parity partners.
One can check for the consequences if one assumes that the opposite-parity nucleonic state is not the N(1535) but another, potentially broad, state with different mass.
The results for the behavior of the scalar condensate $\sigma$ field as function of baryochemical potential for a range of possible masses $m_N(-)$ are shown in Fig. \ref{sigmamnpar}.
One can observe that for a mass above $\approx 1470$ MeV a first-order phase transition occurs in the dense system.

\section{Including Quarks}
\begin{figure}[]
\centerline{\includegraphics[width=8.4cm]{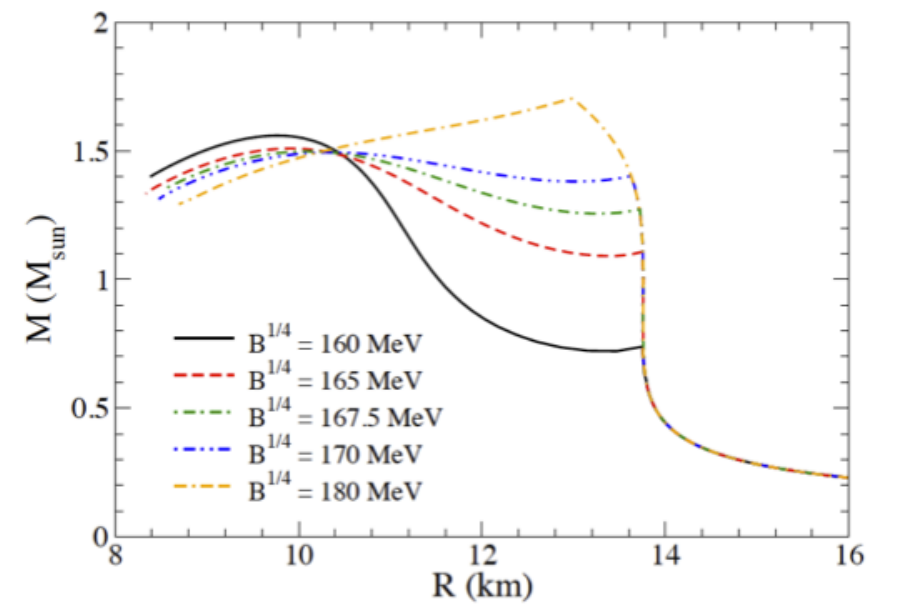}}
\vspace*{5pt}
\caption{Mass-radius diagram for a hybrid star with a MIT bag equation of state for non-interacting quarks. The kink is generated by the onset of the quark phase.}
\label{hybridnoalpha}
\end{figure}

Expanding the investigation of neutron or hyper stars to hybrid stars that also include quarks in the core region often leads to rather small star masses.
As an example this can be seen in Fig. \ref{hybridnoalpha}, where the resulting mass-radius diagrams of hybrid stars are shown. 
Here, a
standard baryonic equation of state G300  \cite{Glendenning} has been connected to a simple equation of state of the MIT bag model. The values of the bag constants have been used in such a way as to make sure that nuclear matter is still the ground state of matter (and not quark matter).
One can observe a reduction of the maximum mass, which is 1.8 solar masses in the case of purely baryonic matter for this specific equation of state. The onset of the quark phase softens the equation of state significantly due to the bag constant, which for large bag constants even leads to a cut-off of the spectrum of stable stars at the onset of the quark phase in the center of the star, rendering hybrid stars unstable.

\begin{figure}[]
\centerline{\includegraphics[width=8.4cm]{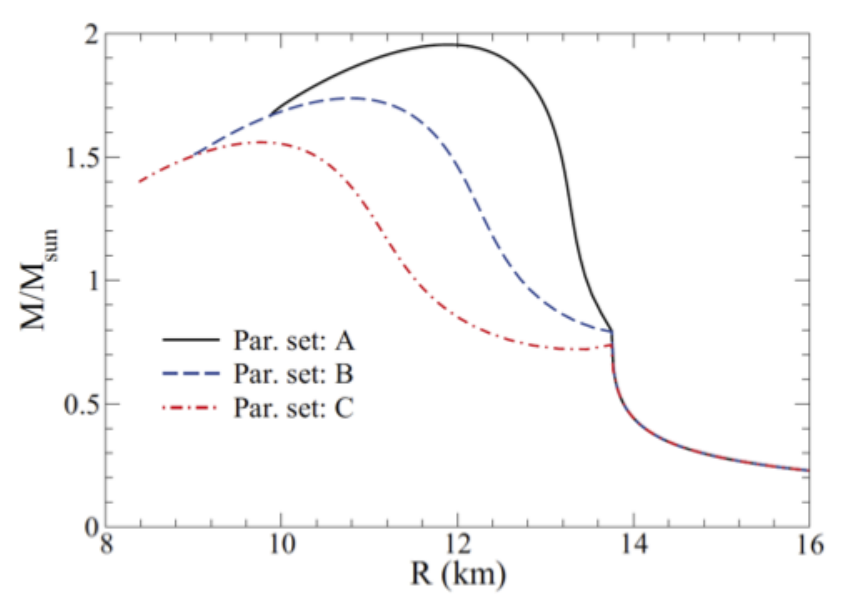}}
\vspace*{5pt}
\caption{Same as Fig. \protect\ref{hybridnoalpha} but including effects of a repulsive quark-quark interaction. The quark interaction strength is increased going from parameter sets C to A. For details on the parameters, see \protect\cite{Negreiros:2010tf}.}
\label{hybridalpha}
\end{figure}

\begin{figure}[]
\centerline{\includegraphics[width=8.4cm]{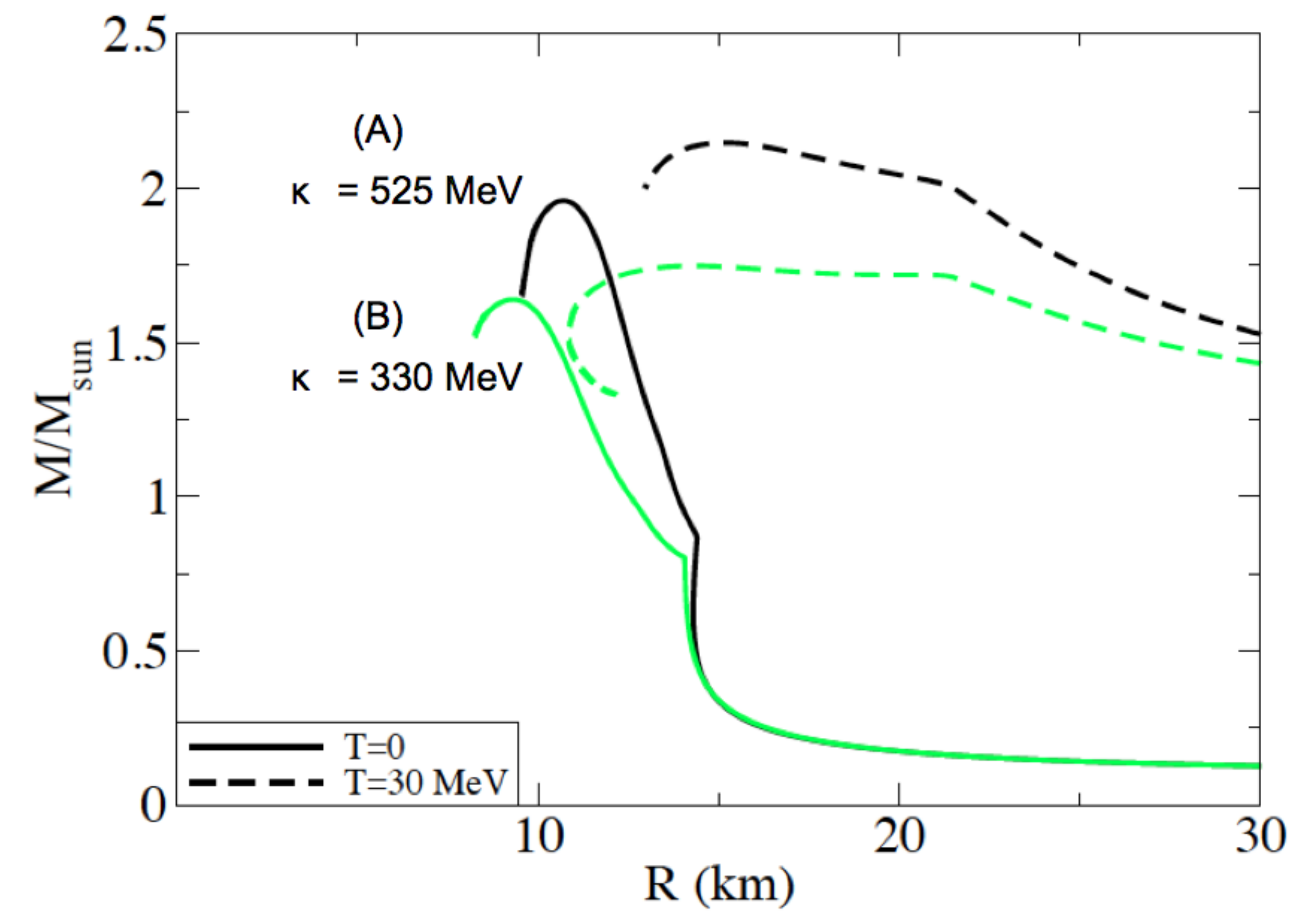}}
\vspace*{5pt}
\caption{Mass-radius diagram of compact stars in the parity-doublet quark-hadron approach. Two parameter sets are shown with different compressibilities of the nuclear ground state.
In addition, the corresponding results for proto-neutron stars with a temperature of 30 MeV are shown \protect\cite{Dexheimer:2012eu}.}
\label{mrpar}
\end{figure}
\begin{figure}[]
\centerline{\includegraphics[width=10.4cm]{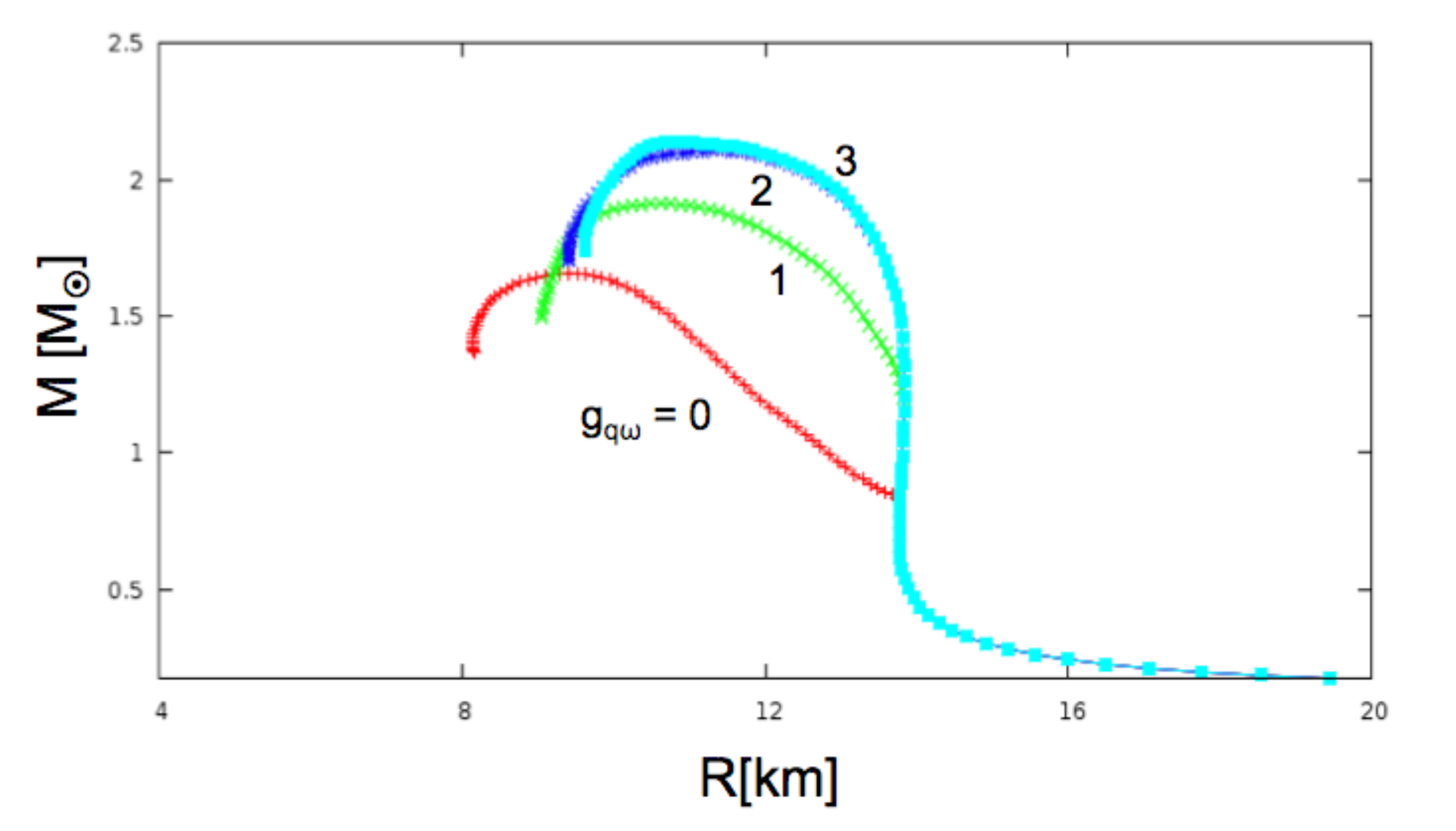}}
\vspace*{5pt}
\caption{Mass-radius diagram for zero-temperature stars in the parity-doublet model including repulsive quark vector-meson interactions. The influence of the quark-vector coupling $g_{q\omega}$ on the attainable maximum masses is shown. }
\label{mrvectorpar}
\end{figure}

In general, the parity-doublet approach leads to an equation of state which tends to be rather stiff. Including relativistic Hartree corrections can ameliorate this behavior to some extent \cite{Dexheimer:2008cv}. Following the original discussion of introducing quarks, the results in Fig. \ref{mrpar} were obtained without quark vector-meson coupling. 
If one includes this coupling one can increase the maximum mass of the star to phenomenologically consistent values also for the case (B) with relatively low compressibility.
This is shown in Fig. \ref{mrvectorpar} for different coupling strengths.

However, including (repulsive) interaction effects in the quark phase can change this figure as originally discussed in 
\cite{nucl-th/0411016}.
The interaction makes the equation of state of the quark phase look more "hadron-like", leading to a smoother transition to a hybrid star.
Fig. \ref{hybridalpha} shows the results of a calculation including quark interactions \cite{Negreiros:2010tf}. As one can observe, increasing the interacting strength leads to higher mass stable hybrid stars.

To overcome the inherent problems of connecting two separate equations of state for hadrons and quarks, as they were discussed in Section 2, 
we have extended our hadronic approach introduced in the previous sections to include quarks in a consistent approach \cite{Dexheimer:2009hi,Steinheimer:2010ib,Dexheimer:2012eu}.
This is done by linearly coupling non-strange (q) and strange (s) quarks to the mean fields as in the case of baryons, Eq. (\ref{efmas}):
\begin{equation}
m_{q}^*=g_{q\sigma}\sigma+\delta m_q~~~
m_{s}^*=g_{s\zeta}\zeta+\delta m_s ~~.
\end{equation}
A coupling of the quarks to the Polyakov loop $\Phi$ is introduced in a similar way as in PNJL quark models\cite{Fukushima:2003fw,Ratti:2005jh}. Their thermal contribution to the grand canonical potential $\Omega$ reads:
\begin{equation}
	\Omega_{q}=-T \sum_{i\in Q}{\frac{\gamma_i}{(2 \pi)^3}\int{d^3k \ln\left(1+\Phi \exp{\frac{E_i^*-\mu_i}{T}}\right)}}
\end{equation}
and a corresponding term for antiquarks. In addition there is a potential for the Polyakov loop, with parameters fitted to quenched lattice QCD results \cite{Ratti:2005jh}. For an in-depth discussion of this approach, 
see \cite{Steinheimer:2010ib}.
Fixing parameters and performing a star calculation leads to results shown in Fig. \ref{mrpar}. The graph shows two fits with compressibilities $\kappa$ = 330 and 525 MeV, respectively with corresponding maximum star masses of 1.75 and 1.96 $M_{\odot}$. One can observe the kink in the curves, signaling the first-order phase transition in this model.

\begin{figure}[]
\centerline{\includegraphics[width=10cm]{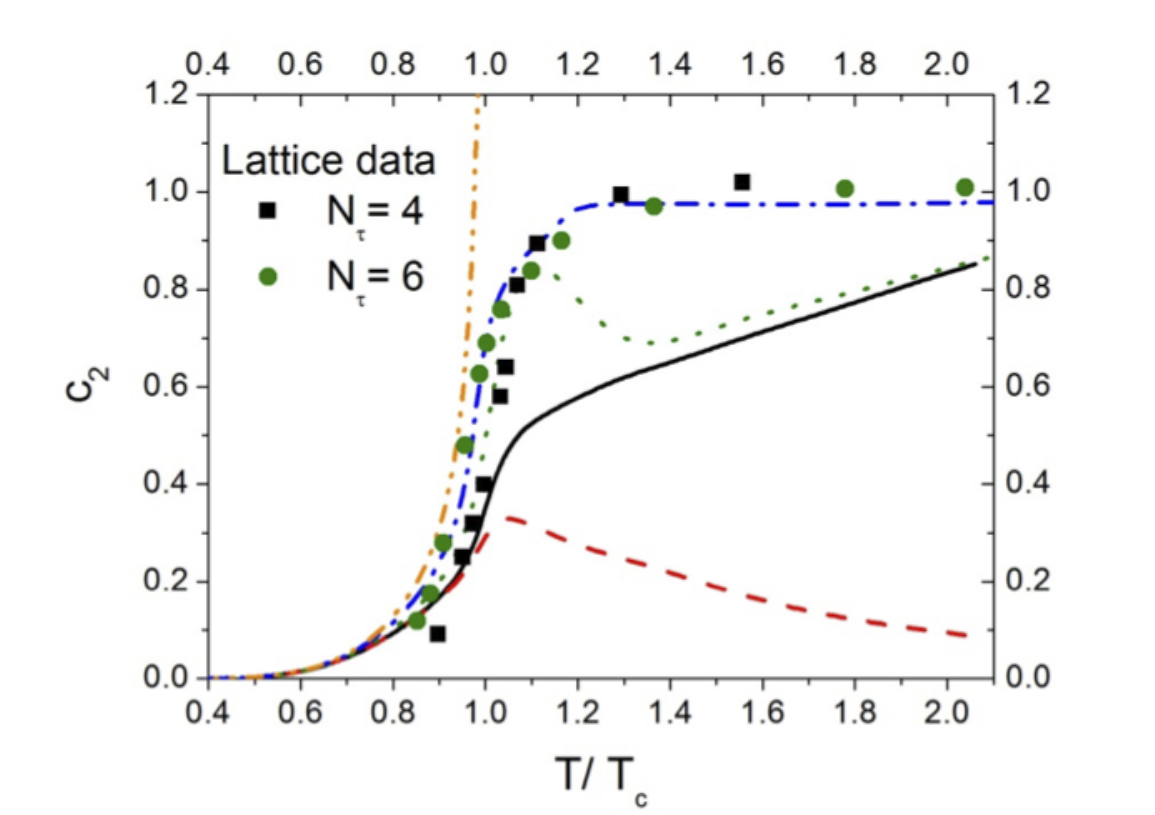}}
\vspace*{5pt}
\caption{Taylor coefficient $c_2$ of the expansion of the pressure with respect to baryon chemical potential.
In addition to lattice QCD results \protect\cite{Cheng:2008zh} the model results for vanishing quark vector interaction (full line) and for a value
of $g_{q\omega} = \frac{1}{3} g_{N\omega}$ (dashed line) are shown. One can observe a significant
deviation from lattice results for non-vanishing vector coupling \protect\cite{arXiv:1005.1176}.}
\label{c2qh}
\end{figure}

Thus, also in such an approach one can achieve a large-mass hybrid star by stiffening the quark equation of state.
However, the introduction of a strong vector interaction term for the quarks leads to severe problems for the description of the system at low chemical potentials.
This can be seen by comparing results of the first non-vanishing Taylor coefficient $c_2$ of the expansion of the pressure $p$ in terms of the quark chemical potential $\mu_q $ over temperature T:
\begin{equation}
c_2(T) = \left. \frac{1}{2} \frac{\partial^2(p(T,\mu_q)/T^4)}{\partial(\mu_q/T)^2}\right|_{\mu_q=0}~~.
\end{equation}
Figure \ref{c2qh} shows a comparison of lattice data for $c_2$ with the model. One can clearly observe a large disagreement of the model results and lattice data for a substantial vector coupling.
These results were obtained in our specific quark-hadron model approach as described above, however the general drop of the curves with increasing quark vector interaction holds for any of the usual NJL, PNJL, or Dyson-Schwinger approaches \cite{arXiv:1005.1176}.
This point remains a major problem for a successful description of massive hybrid stars with a substantial quark core and still needs to be addressed in the future.

\bigskip

\end{document}